\documentclass[prl,showpacs,byrevtex]{revtex4}
\input epsf
\usepackage{bm}
\def\ba{\begin{eqnarray}}
\def\ea{\end{eqnarray}}
\def\br{\begin{array}}
\def\er{\end{array}}
\def\be{\begin{equation}}
\def\ee{\end{equation}}
\def\uoo{c_{13}c_{12}}
\def\uot{c_{13}s_{12}}
\def\uoth{s_{13}}
\def\uto{-c_{23}s_{12}-c_{12}s_{13}s_{23}}
\def\utt{c_{12}c_{23}-s_{12}s_{13}s_{23}}
\def\utth{c_{13}s_{23}}
\def\utho{s_{12}s_{23}-c_{12}s_{13}c_{23}}
\def\utht{-c_{12}s_{23}-c_{23}s_{13}s_{12}}
\def\uthth{c_{13}c_{23}}
\def\be{\begin{equation}}

\def\D31{\Delta m_{31}^2}
\def\D21{\Delta m_{21}^2}
\def\D32{\Delta m_{32}^2}
\def\Dsol{\Delta m_{\odot}^2}
\def\Datm{\Delta m_{atm}^2}
\begin{document}
\title{Threshold Effects on Quasi-degenerate Neutrinos with High-scale Mixing
Unification}
\author{R. N. Mohapatra}
\email{rmohapat@physics.umd.edu}
\affiliation{Department of Physics, University of Maryland, College Park, MD
20742, USA.}
\author{M. K. Parida}
\email{mparida@sancharnet.in}
\affiliation{Department of Physics, North-Eastern Hill University, Shillong
793022,India.}
\author{G. Rajasekaran}
\email{graj@imsc.res.in}
\affiliation{Institute of Mathematical Sciences, Chennai 600113, India.}
\begin{abstract}
We consider threshold effects on neutrino masses and mixings in a
recently proposed model for understanding large solar and
atmospheric mixing angles using radiative magnification for the case of
quasi-degenerate neutrinos. We show that the magnitude of the threshold
effects is sufficient to bring concordance between the predictions of
this model and latest data from ${\rm KamLAND}$ and ${\rm SNO}$ on
observations of neutrino oscillations.
\end{abstract}
\date{\today}
\rightline{NEHU/PHY-MP-01/05}
\maketitle
\par
A major theoretical challenge posed by neutrino observations is how to
understand the large difference between the mixing angles in lepton sector
from that in the quark sector. In a recent paper\cite{parida} we proposed
a mechanism based on the idea that neutrino masses and mixings derived in
a seesaw framework are defined at high scale and, in order to compare them
with experiments, we must extrapolate them to the weak scale using
renormalization group equations\cite{babu}. It had been noted in
ref.\cite{balaji} that if neutrinos are quasi-degenerate and have same CP
property, then small mixings at the seesaw scale can be ``radiatively
magnified'' when extrapolated to the weak scale. The work of
ref.\cite{parida} provided the first realistic model based on this idea.
It started with the hypothesis that at the seesaw scale the
quark and
lepton mixing angles are identical and neutrinos are
quasi-degenerate in mass with same CP property\cite{parida}. These
assumptions were shown to be realizable in a class
of very economical models based on $SU(4)_c$ gauge unification
\cite{ps}. The individual neutrino masses (quasi-degenerate) are the
inputs into the model and the rest are dictated by renormalization group
equations of low energy MSSM. This led  to large mixings in the
solar  and atmospheric neutrino sectors while
maintaining consistency with the small upper limit on the mixing
angle $\theta_{13}$ given by  CHOOZ-Palo-Verde experiments. The precise
values
of the three input neutrino masses are determined
by the present values of the mixing angles. The model has two interesting
predictions: (i) it works only if the  common mass
of quasi-degenerate neutrinos is in the range $0.15$ eV $\le
{\rm m}_i \le 0.65$ eV and (ii) predicts that the lepton mixing angle
$\sin\theta_{13} \le 0.08-0.10$.  The ``large'' value of the common
mass is of great experimental interest since it
falls in the range claimed by the Heidelberg-Moscow experiment\cite{ref2}
and is testable in very near future in all proposed ${\beta\beta}_{0\nu}$
experiments\cite{ref22}. It also has a significant overlap with the
mass range accessible to the KATRIN  experiment  \cite{ref3}. The
predicted value for $\theta_{13}$ is also in the range accessible
by experiments\cite{bnl} being proposed. Therefore this mechanism for
understanding large mixings is an eminently testable proposal.

While the predictions of the model are in good agreement with the
gross features of the experimental data on atmospheric neutrinos
and solar mixing angle, that for $\Delta  m^2_\odot$ is larger
than the latest value derived from the joint analysis of the
 KamLAND and solar neutrino data, which appeared
subsequent to our work. In view of the fact that searches for
neutrinoless double beta decay and $\theta_{13}$ are at the forefront of
neutrino
experimental efforts in the near future and since they also
represent two crucial tests of our model, it is important to address this
discrepancy and search for any other effects within the same model that
can address this problem. In this brief report we point out such an
effect.

In renormalizable field theories, there is indeed another effect that is
relevant
in comparing theory with experiment i.e. the one loop corrections that
involve only the low scale physics. These are known
as threshold effects. In our case it involves the one loop
corrections to the extrapolated effective theory at the weak scale. These
effects for neutrino masses have been discussed in \cite{ref9} and shown
to have a simple form if one assumes that the scalar masses are
universal, as is usually done in MSSM to avoid conflict with flavor
changing neutral current constraints.
They have been found to make significant
contribution on degenerate/quasi-degenerate neutrinos \cite{ref9,ref10}.
In this note we estimate them in the high-scale mixing unification
scenario.

\par
In a typical seesaw model, the neutrino mass matrix at the weak scale can be
written in the flavor basis  as:
\begin{eqnarray}
{\cal M}_\nu(\mu)~=~I(M_R,\mu){\cal M}^0_\nu I^T(M_R,\mu)~+~\Delta {\cal
M}^{th}, \label{eq1}
\end{eqnarray}
where ${\cal M}^0_\nu$ is the neutrino mass matrix at the high
(seesaw) scale and the renormalization effects from the seesaw
scale $M_{\rm R}$ to the weak scale, $\mu$ are represented by a
matrix in flavor space, $I(M_R,\mu)$ and the weak scale threshold
effects are denoted by $\Delta {\cal M}^{th}$. In our model, the
first term in Eq. (1) has already been calculated in
ref.\cite{parida}. Using mass basis with $i,j=1,2,3$ denoting
different mass eigenstates and representing the threshold
corrections through loop factors by $T_{ij}$, the effects on mass
eigenvalues in any model are expressed as\cite{ref9,ref10},
\par
\noindent
\be
m_{ij}=m_i\delta_{ij}+m_i T_{ij}+m_j T_{ji}, \label{eq2}\ee
\par
\noindent It is clear that threshold effects are
significant(negligible) for quasi-degenerate(hierarchical)
neutrinos provided the masses are in the range of interest in our
model (see above). As in \cite{parida}, we  ignore all phases for
the sake of simplicity and   parameterize the real $3 \times 3$
PMNS mixing matrix  as
\par
\noindent
\be U=\left[\br{ccc}
\uoo&\uot&\uoth\\
\uto&\utt&\utth\\
\utho&\utht&\uthth\\
\er\right],\label{eq3}\ee
\par
\noindent
The loop factors in the mass basis can be expressed in terms of those in
the flavor basis,
$T_{\alpha\beta}$($\alpha, \beta= e, \mu, \tau$) leading to (for
quasi-degenerate neutrinos)
\par
\noindent
\be m_{ij}=m_i \delta_{ij}+2m \sum_{\alpha,\beta} T_{\alpha\beta}
~U_{\alpha\rm i} U_{\beta \rm j}. \label{eq4}\ee
\par
\noindent where ${\rm m}$ is the common mass of quasi-degenerate
neutrinos. In the MSSM flavor violation through threshold
corrections is constrained to its minimal value through the
diagonal structure of loop factors, $T_{\alpha \beta}=T_{\alpha}
\delta_{\alpha\beta}$. Experimental data show that atmospheric
neutrino mixing angle is close to its maximal value ,
$\theta_{\rm atm}=\theta_{23} \simeq \pi/4$ and the reactor
mixing angle  $\theta_{\rm CHOOZ}=\theta_{13} < {10}^o$ whereas
the solar neutrino mixing angle deviates from its maximal value,
$\theta_{sol}$=$\theta_{12} \simeq {31}^o-{34.5}^o$. Then any
result derived under the limiting case $\theta_{atm}$=$\pi/4$ and
$\theta_{\rm CHOOZ}$=0 is expected to hold in the actual case
with a very good approximation. Using  eqs.(1)-(3) and
$\theta_{23}$=$\pi/4$, $\theta_{13} \rightarrow 0$  the following
relations are derived which are valid in any model under the
assumption of minimal flavor violation:
\par
\noindent
\ba (\Delta m_{21}^2)_{th}
=4\rm m^2\cos 2\theta_{12}[-T_e+(T_\mu+T_\tau)/2], \label{eq5}\\
(\D32)_{th}
=4\rm m^2\sin^2\theta_{12}[-T_e+(T_\mu+T_\tau)/2], \label{eq6}\\
(\Delta m_{31}^2)_{th}
=4\rm m^2\cos^2\theta_{12}[-T_e+(T_\mu+T_\tau)/2], \label{eq7}\ea
\par
\noindent
These equations in turn establish interrelations among threshold
corrections to the
three mass squared differences:
\par
\noindent
\ba (\Delta m_{21}^2)_{th}=(\cot^2\theta_{12}-1)(\D32)_{th},
\label{eq8}\\
=(1-\tan^2\theta_{12})(\Delta m_{31}^2)_{th},\label{eq9}\\
(\Delta m_{32}^2)_{th}=\tan^2\theta_{12}(\Delta m_{31}^2)_{th},
\label{eq10}\\
(\Delta m_{21}^2 )_{th}-(\Delta m_{31}^2)_{th}+
(\Delta m_{32}^2)_{th}=0, \label{eq11}\ea
\par
\noindent
Eq. (11) says that the tree-level relation among the mass-squared
differences is also true at one-loop level as is evident from Eqs.(8)-(10).
\par
It is also clear from Eq. (8) that for allowed values of $\theta_{\odot}
={32.6}^o\pm {1.6}^o$,
 $(\Delta m_{21}^2)_{\rm th}=($5.5-1.2$)(\Delta m_{32}^2)_{\rm
th}$. Therefore any attempt to match the experimentally observed values of
$\Delta m_{atm}^2 \simeq 2\times 10^{-3}$ eV$^2$ through $(\Delta
m_{32}^2)_{th}$ and $\Delta m_{\odot}^2$ through $(\Delta
m_{21}^2)_{\rm th}$ would lead to the prediction in the solar neutrino
sector $\Delta m_{\odot}^2 \approx 11.0\times 10^{-3}$
eV$^2$ -$2.4\times
10^{-3}$ eV$^2$.
Since these are
at least two orders larger than the experimentally permissible range,
we conclude that the low-energy threshold effects alone in any model
can not accommodate
the observed mass squared differences for solar and the atmospheric neutrinos.

\par
Coming to the case of quasi-degenerate neutrino model of
ref.\cite{parida}, first
point to note is that, there is more to the mass difference squares in
this model than
the threshold corrections. Therefore even though the
 $(\Delta m^2)_{th}$'s satisfy Eq. (5-7) they do not a priori lead
to any contradiction with observations.
For this case, it was shown in ref.\cite{parida} that
for a wide range of GUT-seesaw scales, $M_{\rm R}$=
$10^{11}$ GeV - $10^{18}$ GeV, one has the following predictions of the
neutrino oscillation parameters:
\par
\noindent
\ba (\sin\theta_{12})_{\rm RG}&=&0.42-0.62,~(\sin\theta_{23})_{\rm
RG}=0.660-0.707, \nonumber\\
(\sin\theta_{13})_{\rm RG}&=&0.08-0.10,~(\Delta m_{32}^2)_{\rm
RG}=(1.18-4.6)\times 10^{-3} eV^2,\nonumber\\
(\Delta m_{21}^2)_{\rm RG}&=&(1.20-6.0)\times 10^{-4} eV^2. \label{eq12}\ea
From the global fits of the solar neutrino data within $1\sigma$ limit
 $\Delta m_{\odot}^2$=$(5-8)\times 10^{-5}$ eV$^2$. Thus, the predicted
values of this quantity in the high-scale mixing unification
model are at least $3\sigma-4\sigma$ larger  while all other
physical parameters are in good agreement with the experimental
values. We show that quite reasonable and plausible threshold
corrections due to super-partners at the electro-weak scale with
minimal flavor violation which were ignored earlier are
sufficient to account for this discrepancy. Other physical
parameters undergo quite small or negligible threshold
corrections while maintaining their agreement with the
experimental data. Including threshold corrections along with the
RG-evolution effects we define,
\par
\noindent
\be\Delta m_{ij}^2=(\Delta m_{ij}^2)_{\rm RG}+
(\Delta m_{ij}^2)_{\rm th} \label{eq13}\ee
\par
\noindent
where , as stated through (12), the RG-evolution effects from $M_{\rm R}$
to $M_{\rm Z}$ have been already computed in \cite{parida}. ~We find that
the
simple form of the one-loop SUSY threshold effect with wino/slepton exchange
in the loop is sufficient to give the desired correction. These
effects due to gaugino/slepton exchange are evaluated
using\cite{ref10} and we find
\par
\noindent
\ba T_{\alpha}=(g^2/ {32\pi^2})[(x_{\mu}^2-x_{\alpha}^2)/
(y_{\mu}y_{\alpha})+((y_{\alpha}^2-1)/ y_{\alpha}^2)
ln(x_{\alpha}^2)- ((y_{\mu}^2-1)/ y_{\mu}^2)ln(x_{\mu}^2)] \label{eq14}\ea
\par
\noindent
where $y_{\alpha}$=$1-x_{\alpha}^2$, $x_{\alpha}$ = $M_{\alpha}/M_{\tilde w}$,
$M_{\alpha}$ = charged slepton mass, and $M_{\tilde w}$ =  wino mass.
The  loop-factor has been defined to give $T_{\mu}$ = 0 without any
loss of
generality\cite{ref10}. For several allowed mass ratios
$M_{\tilde {\rm e}}/M_{{\tilde \mu},{\tilde \tau}}$
in the MSSM we estimate threshold corrections
$(\Delta m_{32}^2)_{th}$ and  $(\Delta m_{21}^2)_{th}$ as shown in Table 1.
We find that for inverted hierarchy in the charged-slepton sector with
$M_{\tilde {\rm e}}/M_{{\tilde \mu},{\tilde \tau}}$=$1.3-2.5$ the
corrections are sufficient for agreement with the solar neutrino data
within $1\sigma$ limit. The correction to
$(\Delta m_{32}^2)_{th}$ is also negative but about one-order smaller
than the corresponding RG contributions. The threshold corrections have
negligible effects on the mixing angles obtained by RG-evolution. We give
our detailed results in Table I.

 We also observe that the inverted hierarchical form of neutrino mass eigenvalues,
namely, $m_2\agt m_1 \agt m_3$ is not allowed in the radiative
magnification  scenario\cite{parida}. We have found that in the
MSSM with low-energy SUSY, the $\beta$-function coefficients of
the three mass eigen values are positive near the see-saw scales
with the third $\beta$-function coefficient being significantly
larger  compared to the first two. This is the basic reason why
all the mass eigen values decrease although with different rates
finally approaching a common value at the lowest-SUSY scale where
magnification takes place. When applied to inverted hierarchical
~pattern,~the mass eigenvalues move away from one another instead
of approaching a common value at the SUSY scale.
\par
In summary, we have calculated the threshold corrections to the
neutrino masses and mixings in a high scale mixing unification model
proposed recently to understand large mixings. We find that this
substantially improves agreement of the model with the latest observations
for $\Delta m^2_\odot$ as well as other oscillation observables.  We have
also argued on general grounds that both $\Dsol$ and $\Datm$ can not  be
obtained by low-energy threshold effects alone on quasi-degenerate neutrinos.

\begin{table*}
\caption{Radiative magnification to bilarge mixings and threshold
correction effects
at low energies for input values of ${\rm m}^0_{\rm i}$($i=1,2,3$),
 $\sin\theta_{23}^0=0.038$,~$\sin\theta_{13}^0=0.0025$, ~and
$\sin\theta_{12}^0=0.22$ at the high scale $M_R$.}
\begin{ruledtabular}
\begin{tabular}{lccccc}\hline
$M_{\rm R}$(GeV)&$10^{11}$&$10^{13}$&$10^{13}$&$10^{15}$&$2\times10^{18}$\\

$m_1^0$(eV)&0.4083&0.5170&0.6168&0.3980&0.5160\\

$m_2^0$(eV)&0.4100&0.5200&0.6200&0.4000&0.5200\\

$m_3^0$(eV)&0.4510&0.5910&0.7050&0.4730&0.6680\\

$m_1$(eV)&0.2723&0.3107&0.3719&0.2093&0.1714\\

$m_2$(eV)&0.2726&0.3122&0.3723&0.2098&0.1718\\

$m_3$(eV)&0.2745&0.3152&0.3759&0.2124&0.1750\\

$(\Delta m^2_{21})_{\rm RG}$(eV$^2$)&$1.6\times 10^{-4}$&$3.0\times 10^{-4}$&

$3.5\times 10^{-4}$&$2.0\times 10^{-4}$&$1.36\times 10^{-4}$\\

$(\D32)_{\rm RG}$(eV$^2$)&$1.18\times 10^{-3}$&$1.8\times 10^{-3}$&

$2.6\times 10^{-3}$&$1.15\times 10^{-3}$&$1.18\times 10^{-3}$\\

$M_{\tilde e}/M_{\tilde \mu, \tilde \tau}$&1.5&1.9&2.3&1.9&1.5\\

$(\Delta m^2_{21})_{\rm th}$(eV$^2$)&$-0.85\times 10^{-4}$&$-2.3\times

10^{-4}$&$-2.8\times 10^{-4}$&$-1.25\times 10^{-4}$&$-0.57\times 10^{-4}$\\

$(\D32)_{\rm th}$(eV$^2$)&$-0.60\times 10^{-4}$&$-1.3\times 10^{-4}$&

$-3.6\times 10^{-4}$&$-0.48\times 10^{-4}$&$-0.15\times 10^{-4}$\\

$\Delta m^2_{\odot}$(eV$^2$)&$7.5\times 10^{-5}$&$7.0\times 10^{-5}$&

$7.0\times 10^{-5}$&$7.0\times 10^{-5}$&$7.8\times 10^{-5}$\\

$\Delta m^2_{\rm atm}$(eV$^2$)&$1.12\times 10^{-3}$&$1.7\times 10^{-3}$&

$2.2\times 10^{-3}$&$1.1\times 10^{-3}$&$1.17\times 10^{-3}$\\

$\sin\theta_{12}$&0.550&0.520&0.550&0.533&0.522\\

$\sin\theta_{23}$&0.700&0.707&0.690&0.695&0.696\\

$\sin\theta_{13}$&0.100&0.104&0.097&0.100&0.098\\\hline
\end{tabular}
\end{ruledtabular}
\label{tab1}
\end{table*}
\begin{acknowledgments}
This work was initiated at the Eighth Workshop on High-Energy Physics
Phenomenology(WHEPP8), Mumbai, 2004.
M.K.P. thanks the Institute of Mathematical Sciences, Chennai for Senior
Associateship.
The work of R.N.M is supported by the NSF grant No.~PHY-0354401. The work
of G.R. is supported by the DAE-BRNS Senior Scientist Scheme  of the
Govt. of India.
\end{acknowledgments}

\end{document}